\documentclass{PoS}

\title{Vertical Array in Space for Horizontal Air-Showers}

\ShortTitle{Vertical Array in Space}

\author{\speaker{Daniele Fargion}\thanks{The article is devoted to the memory of  astronaut Ilan and his son, pilot Asaf  Ramon, lost deep in the sky }\\
        Author affiliation: Physics Department, Rome University 1 and INFN rome1, Ple. A. Moro 2, 00185, Rome, Italy\\
        E-mail: \email{daniele.fargion@roma1.infn.it}}

\abstract{
          Since a century cosmic rays are based on direct cosmic particle detection in space (below PeV)  or on secondary downward vertical airshowers (above TeVs). We consider  the guaranteed physics of horizontal (hadron) air-showers, HAS,  developing at high ($30-40$) km altitudes, above and below these energy windows. Their morphology and information traces are different from vertical ones. Hundreds of km long HAS are often split by geomagnetic fields in a long (fan-like) showering with a twin spiral tail.  The horizontal fan-like airshowers are really  tangent and horizontal only at North and South poles. At different latitude their showering plane are turned and inclined by geo-magnetic fields. In particular at magnetic equator such tangent horizontal East-West airshowers are bent and developed into a vertical fan air-shower, easily  detectable by a vertical array detector (hanging elements by gravity). Such \emph{medusa } arrays maybe  composed  by inflated floating balloons chains. The light gas float and it acts as an calorimeter for the particles, while it partially  sustains the detector weight.  Vertically hanging chains  as well as  rubber doughnut balloons ( whose  interior may record Cherenkov lights)  reveal bundles of crossing electron pairs.  Even in space orbit such vertical array may hang by tiny tidal forces within huge balloon arrays, while brought in locus  by an extendable measure tape. Possibly located around Space Station in synergy with future AMS-particle detector.  Such an array  maybe loaded at best and cheapest prototype  in common balloons tracing upward and tangent hadron air-showers from terrestrial atmosphere edge. These array structure may reveal the split shower signature. Offering a way to disentangle better their shower origination, energy and interaction point. Better revealing the composition nature. Just beyond the Earth horizons there are exciting, but rarer, sources of upward airshowers: the new UHE Tau Air-showers astronomy originated within Earth by EeVs  tau  neutrino signals skimming the soil and forming UHE Tau, decaying later in flight. Therefore Horizontal airshowers at equator may show the hadron horizontal twin split nature, its composition as well as a very first expected  UHE Neutrino astronomy.\ }

\FullConference{European Physical Society Europhysics Conference on High Energy Physics,
EPS-HEP 2009,\\
		 July 16 - 22 2009\\
		 Krakow, Poland}

\begin{document}

\section{Vertical Airshower versus Horizontal ones}
\begin{figure}[htb]
\begin{center}
\includegraphics[width=12cm]{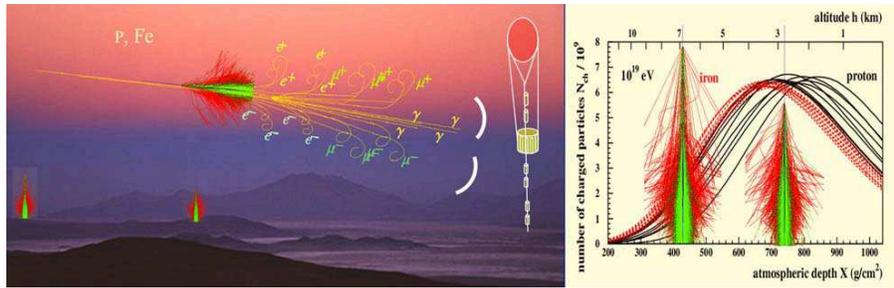}
\caption{Horizontal Air-Showers versus Vertical ones; the composition nature of the primary defines the place of the interaction. The
spitting of the horizontal airshower maybe tested by different element spread vertically as in the tail array shown in figure}
\label{vertical}
\end{center}
\end{figure}

The atmosphere above us at sea level is comparable to $10$ meter water slant depth, defending our life  from cosmic ray radioactivity hazard.
This healthy  atmosphere screen does also blur the cosmic ray propagation and its pair-production tail. Unfortunately in addition galactic and cosmic bending makes Cosmic Rays mostly smeared out hiding any Cosmic Ray Astronomy. (UHECR nevertheless might already or soon offer, somehow, a nearby GZK Astronomy \cite{Auger07},\cite{Farg08a}). Therefore the vertical air-showering pair-production is the main tool in UHECR: it does form and expand (mostly by Coulomb scattering)  into a confused conical Christmas tree. Because of it we loose information on the primary interaction altitude and its probable composition nature. Moreover vertical airshowers (at energy below EeV)  have lost most of abundant electromagnetic  tails, absorbed along the flight, once at sea  level; only rarer muons (the penetrating  lepton secondaries) are surviving at the sea level. At highest altitudes vertical airshower are impossible because of the negligible residual slant depth. Nevertheless at high altitude horizontal airshowers are well possible because they are crossing a longer transverse atmosphere layer. Above tens ( nearly $40$)  km. the horizontal and even upward hadronic airshowers may explode and expand into a diluted atmosphere. Upward airshowers from much below are crossing a too deep atmosphere and they are suppressed. Therefore at each direction and each energy the HAS has a fine tuned ideal observer. Viceversa at any high altitude an observer may select and tune its detector to desired energy windows in different zenith angles. Such a "clever" filter is missing in most ground vertical air-showering. Moreover the small air density at high altitudes allow to develop a more collimated airshower, more sensible  to geo-magnetic Lorentz bending. Because at Geo-equator fields Fig\ref{GeoEq}, this bending deflect only vertically (along East West)  ideally for vertical array hanging in space  Fig.\ref{vertical}. Indeed the richest  electron pairs tails maybe even split in twin lobes. Also gamma tail interacting at later air target slant-depth may recreate additional electron pairs and tails.  Therefore the HAS shower exhibits  twice a twin spiral tail.  Finally, even muon pairs are partially bent; but being more energetic (otherwise decaying in flight) and being less abundant they are less separated and diluted. The overall consequence is that horizontal airshowers are much longer (20-40 times or hundreds km.) than vertical ones, opening  into a huge flat fan split air-shower. This spectacular bi-modular shape could not be easily observed  at the ground. However a very spectacular inclined twin split airshower event have been predicted and partially observed in AUGER: both its blazing Cherenkov and its  lateral fluorescence lights as well as its  partial oval muons bundle at ground have been revealed \cite{Farg08a},\cite{Farg08b}. In a very near future (just this year or next) we predict that the horizontal airshower HAS split in West-East directions might be observed by HEAT fluorescence detector in AUGER, at their top telescope edges \cite{Farg09}, see insert Fig\ref{Tau}.
\section{UHECR and Tau Astronomy: Rate following AUGER Cen A discover}
\begin{figure}[htb]
\begin{center}
\includegraphics[width=8cm]{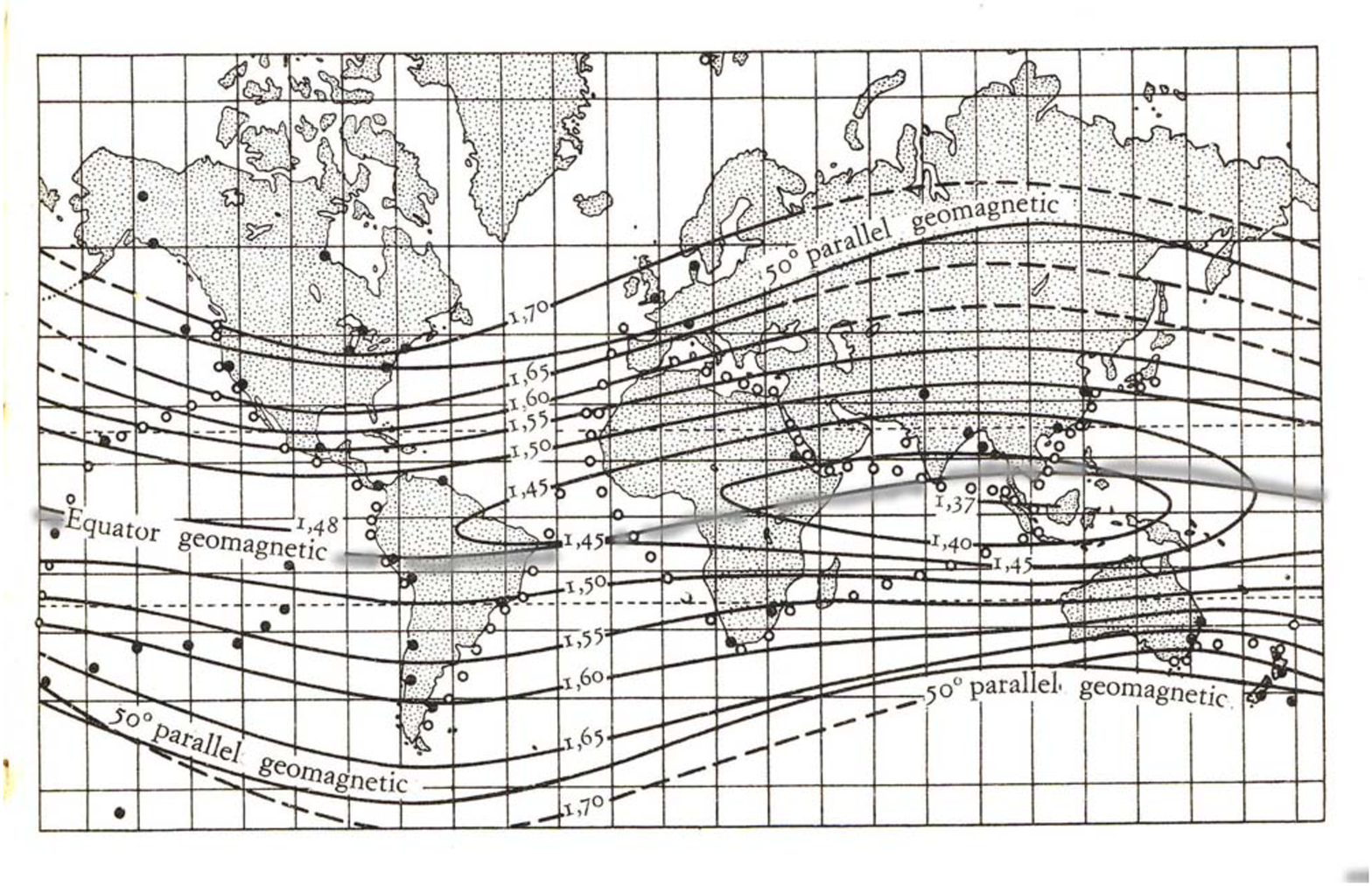}\caption{The Geomagnetic map and the equatorial line where HAS along East-West become vertical ones.}
\label{GeoEq}
\includegraphics[width=10cm]{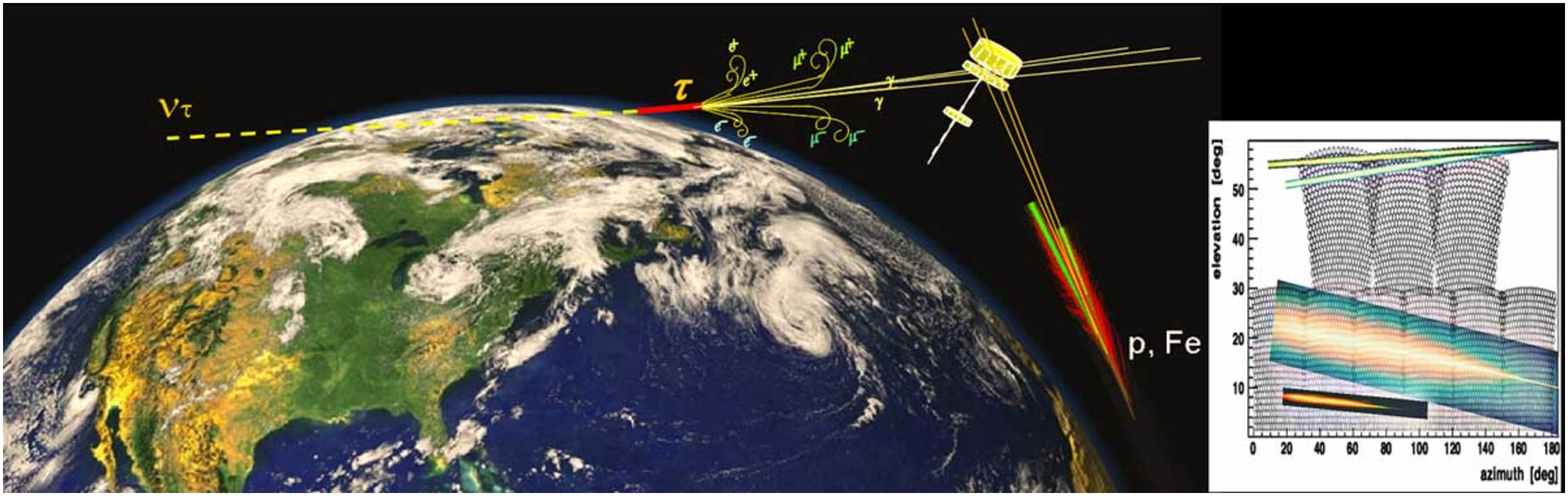}
\caption{A space station model array made by different inflated element at large (km) distance. Such an array will be able to disentangle mainly a prompt neutron burst, originated at atmosphere last layer, followed by and secondaries nucleons and electron pairs at hundreds GeV. These HAS tangent to geo-magnetic field along North-South  directions may beam the shower almost unbent; their GeVs or tens GeVs electron pairs may also reach with an abundant flow the detector almost un-deflected. The night time shower signals, made by  beamed Cherenkov lights of electron secondaries may also be revealed by the complementary detectors with an eye-fish $360^o$  optics pointing on the Earth edges. Such a wide array will look for an area opposite to EUSO-JEM project  horizons. It will be at best located along and around the International Space Station. In possible synergy with future AMS:  The outcome Tau Airshower at few EeV energies might be detectable just below the Earth edge ($1^o$ depart from hadronic HAS horizons ones), (if the UHECR are UHE nucleons degraded by GZK cut-off) \cite{Auger07}. Otherwise, if UHECR are lightest nuclei \cite{Farg08a},\cite{Farg09}, the signal will be quite suppressed at EeV , enhanced at tens PeV, better detectable by deep valley array detectors \cite{Farg97},\cite{Farg09}.   }
\label{Tau}
\end{center}
\end{figure}

The need to use an array in flight to disentangle UHECR composition is compelling. For an example the PeV event rate R observed in air, at a nominal distance from the core of half a km,  from a few hundreds meter balloon array-length, each element made by a meter square area, could be as large as $R\simeq 1-0.1$ Hz. At such altitudes there are important signal correlated to the electron pair showering by geo-synchrotron radio signals. The possibility to reveal an  airshower at horizons or below the horizons at $28$ km require an angular resolution of $\simeq 6-7^o$. Therefore an eventual GZK Tau airshower at EeV maybe well disentangled. At lower altitudes there are also the possible energy windows at $10^{16}- 10^{17}$ eV for UHE tau neutrino possibly relic of UHECR fragmentation. Their arrival will be more easier to disentangle. In some model related to UHECR composition and Cen A event map, this is the case \cite{Farg09}. Therefore while we predict an event or two split at Ten-EeV observable by AUGER-HEAT  telescopes ( see insert in Fig.\ref{Tau}) we predict the abundant detection in balloons array in a very simple way of HAS showers whose bi-modal bumps could be a novel road to sky Cosmic Ray Spectroscopy.
\begin{figure}[htb]
\begin{center}
\includegraphics[width=14cm]{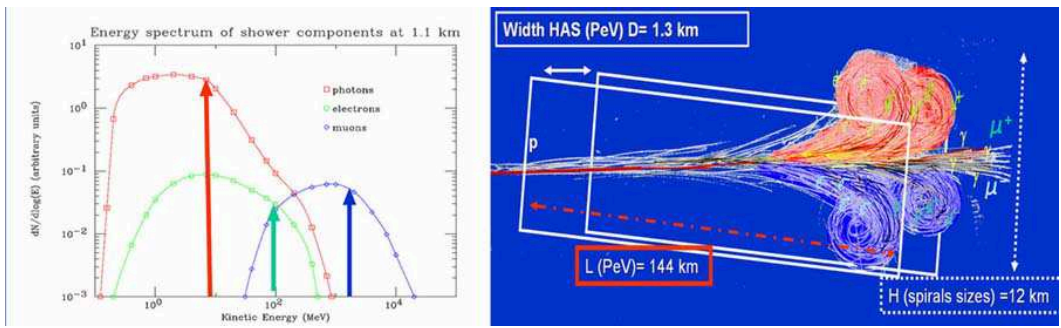}
\caption{A schematic picture of a High Altitude Airshower, HAS; the energy assumed is PeV energy range in horizontal airshower at 28 km altitude. From the approximated energies above each component spectra reflects into a characteristic spiral rings. The wider rings are for more energetic first generation pairs while the smaller ones are for lower energies electron  pairs (a second generation of electron pairs made by photon interacting in diluted air nearly $5-6$ km later a first electron pairs). The muon are more energetic and therefore less bent. At geomagnetic equator the air-shower East-West are spread vertically. Its twin beam and the zenith angle might offer a tool for disentangle the interaction place and the composition. The height H grows in a first approximation, quadratically with the distance L. The width D grows linearly with the distance L. The size length L for a full airshower development is inversely proportional to the air density and proportional to the slat  depth  or  Airshower logarithm energy . }
\label{pulse}
\end{center}
\end{figure}

\end{document}